\documentclass[a4paper]{revtex4}
\usepackage[top=1in, bottom=1in, left=1in, right=1in]{geometry}
\usepackage{amsmath}
\usepackage{amssymb}
\usepackage{amsfonts}
\usepackage{graphicx,bm}
\usepackage{dcolumn}
\usepackage{epsfig}
\usepackage[colorlinks,linkcolor=blue,citecolor=blue,urlcolor=blue]{hyperref}

\def \lleq {\lower0.9ex\hbox{ $\buildrel < \over \sim$} ~}
\def \ggeq {\lower0.9ex\hbox{ $\buildrel > \over \sim$} ~}

\def \beq  {\begin{equation}}
\def \eeq  {\end{equation}}
\def \ber  {\begin{eqnarray}}
\def \eer  {\end{eqnarray}}

\newcommand{\be}{\begin{equation}}
\newcommand{\ee}{\end{equation}}
\newcommand{\ba}{\begin{eqnarray}}
\newcommand{\ea}{\end{eqnarray}}
\newcommand{\bea}{\begin{eqnarray*}}
\newcommand{\eea}{\end{eqnarray*}}

 \newcommand{\bq}{\begin{equation}}
 \newcommand{\eq}{\end{equation}}
 \newcommand{\bqn}{\begin{eqnarray}}
 \newcommand{\eqn}{\end{eqnarray}}
 \newcommand{\nb}{\nonumber}
\begin{document}

\title{Dynamics of coupled phantom and tachyon fields}
\author{M. Shahalam$^1$\thanks{%
E-mail address: shahalam@zjut.edu.cn},  S. D. Pathak$^2$\thanks{%
E-mail address: shankar@sdu.edu.cn}, Shiyuan Li$^2$\thanks{%
E-mail address: lishy@sdu.edu.cn}, R. Myrzakulov$^3$\thanks{%
E-mail address: rmyrzakulov@gmail.com}, Anzhong Wang$^{1,4}$\thanks{%
E-mail address: Anzhong_Wang@baylor.edu}}
\affiliation{$^{1}$Institute for Advanced Physics $\&$ Mathematics, Zhejiang University of Technology, Hangzhou, China\\
$^2$School of Physics, Shandong University, Jinan, China \\
$^3$Eurasian International Center for Theoretical Physics, Department of General and Theoretical Physics, Eurasian National
University, Astana, Kazakhstan\\
$^4$GCAP-CASPER, Department of Physics, Baylor University, Waco, Texas, USA}
\begin{abstract}
In this paper, we apply the dynamical analysis to a coupled phantom field with scaling potential taking particular forms of the coupling (linear and combination of linear), and present phase space analysis. We investigate if there exist late time accelerated scaling attractor that has the ratio of dark energy and dark matter densities of the order one. We observe that the scrutinized couplings cannot alleviate the coincidence problem, however acquire stable late time accelerated solutions. We also discuss coupled tachyon field with inverse square potential assuming linear coupling.
\end{abstract}

\maketitle

\section{Introduction}
\label{sec:intro}
The late time cosmic acceleration is revealed by various observations \cite{planck, perlmutter, riess, Spergel, Komatsu}. A  substantial efforts were put by number of authors to explore the cause of cosmic acceleration, by introducing a new player with negative pressure termed as dark energy (DE) \cite{review}. Apart from dark energy, there are other theoretical models, such as void models and Back-reaction,  which all provide  late time cosmic acceleration  \cite{tomita}. 

The simplest candidate of DE is the cosmological constant $\Lambda$ with the equation of state $w=-1$. However, it suffers two severe problems, such as cosmological constant (fine tuning) and coincidence problems \cite{derev}. Though $\Lambda$CDM model is supported by the  present observations, yet it has no satisfactory argument for fine tuning  and  coincidence problems; why the vacuum energy is so small?  why the densities of DE and dark matter (DM) are nearly equal at present, while their time evolution is much different? Therefore, one can explore the dynamical DE models that can fit into the observations. Such models have been studied in the past few decades \cite{ratra, caldwell, alam, alam2, sd, quint, phant,Boisseau:2000pr, ArmendarizPicon:2000ah, saibal, ira, udm, dutta}.

The simplest models of dynamical DE are scalar fields, dubbed as ``quintessence''. If the quintessence is coupled with the DM, then one can get similar energy densities in the dark sector at present. A conclusive way is if the DE models have $\Omega_{DE}/ \Omega_{DM}$ of the order 1 and an accelerated scaling attractor solution, then the coincidence problem can be alleviated. Therefore, to sort out the coincidence problem, the interaction of DE with DM is one novel approach.

It has been found that  the form of dark energy that dominates  the present Universe could be a phantom energy, quintessence or cosmological constant. The available cosmological data do not fix a microscopic theory of dark energy.  But the overall uncertainty is reflected by the existence of  various phenomenological models.   To reduce the number of models one way is to consider only the ones that do not violate any of the fundamental theories.  The number can be further reduced by testing  the models against the cosmological data. Phantom field  can be  a source of dark energy and may  arise from higher order theories of gravity, for example, the Brans$-$Dicke and non-minimally coupled scalar field theories \cite{bd,amend}. Recently, the dynamics of a coupled phantom field with dark matter has been discussed  \cite{guo}. To solve the long standing coincidence problem, we consider scalar fields (specifically phantom and tachyon) as a dynamical dark energy interacting with dark matter by transferring energy between the two dark components. For an exponential potential, the quantity $\lambda=- V'/\kappa V$, which corresponds to the relative slope of the potential, is constant. Therefore, it is easy to study the stability of the stationary points in the phase space \cite{dw}. 

In the literature, it has been proposed that rolling tachyon condensates, in a class of string theories, may have important cosmological outcomes. Ashoke Sen \cite{sen} has shown that the decay of D-branes generates a pressure-less gas having definite energy density that looks like classical dust. The equation of state of a rolling tachyon lies between 0 and $-$1 \cite{gibbons,paddy}. In this case, we consider inverse square potential for which $\lambda$ is constant, an analogue of exponential potential for standard scalar field. Coupling with matter might lead to late time acceleration. Tachyon field also has implication for inflation, namely, tensor to scalar ratio is very low in this case.

A dynamical system plays a central role in the understanding of the asymptotic behavior of the cosmological models and belongs to the class of autonomous systems \cite{dw}. For an autonomous system, the dimensionless set of variables are chosen due to a number of reasons.
\\
(a) These variables give rise to a bounded dynamical system.\\
(b) They are well-behaved and regularly have a direct physical interpretation.\\
(c) Due to a symmetry in the equations, the number of equations can be reduced and then resulting simplified system is investigated. The brief analysis of the dynamical system is given in Appendix.

In this letter,  we investigate the stationary points and their stability for coupled phantom and tachyon fields. We apply dynamical system analysis to study the asymptotic behavior of the cosmological models mentioned above. We consider the forms of coupling that is proportional to the time derivative of their energy densities. The
different forms of coupling have been studied  in \cite{alamepjc, Boehmer, Cen, Malik, Zia, a35}. There also exist to studies of the models without such particular forms of coupling \cite{a36}. The rest of the paper is organized as follows: In Sect. \ref{sec:cpd} we discuss the coupled phantom dynamics and construct the autonomous system which is useful for phase space analysis. In Sect. \ref{sec:stat} we study phase space trajectories, and obtain stationary points and their stabilities for different forms of coupling. The stationary points and their stabilities of a tachyon field with the coupling $Q=\beta \dot{\rho_{\phi}}$ is discussed in Sect. \ref{sec:ctd}. We summarize our results in Sect. \ref{sec:conc}.
\section{Coupled phantom dynamics}
\label{sec:cpd}
In a spatially flat Universe, we consider two components, namely phantom field and matter (Baryonic+DM). The energy density of each component may not be conserved, although the total energy density of the Universe is. Therefore, the conservation laws  of energy can be written as   
\begin{eqnarray} 
\dot{\rho}_{m}+3H(\rho_{m}+p_m)=Q, \nonumber \\
\dot{\rho}_{\phi}+3H(\rho_{\phi}+p_{\phi})=-Q,    \nonumber \\    
\dot{\rho}_{tot}+3H(\rho_{tot}+p_{tot})=0, 
\label{conser}
\end{eqnarray}
where $\rho_{tot}=\rho_{\phi}+\rho_{m}$ and $p_{tot}=p_{\phi}+p_{m}$, and $\rho_m$, $\rho_{\phi}$, $p_m$ and $p_{\phi}$ are the energy densities and pressures of matter (dust) and phantom filed,  respectively. The coupling is through the function $Q$, and $H$ denotes the Hubble parameter.

The flow of energy between two components depends on the sign of $Q$. If $Q \ > 0$, the transfer of energy takes place from phantom to matter,  whereas for $Q \ < 0$ it occurs from matter to phantom.
At the present, several forms of $Q$ have been investigated  \cite{y1, tapan, z1,z2,z3, chena, Maartensa,nicola,newref}. Following equation (\ref{conser}), it is clear that $Q$ should be a function of $H$, $\rho_m$ and $\rho_{\phi}$,
\begin{eqnarray}
Q= Q(H, \rho_m, \rho_{\phi}).
\label{Q0}
\end{eqnarray}
In this Letter, we consider three particular forms of $Q$: $\alpha\dot{\rho}_{{m}}$, $\beta\dot{\rho_{\phi}}$ and $\sigma (\dot{\rho}_{{m}}+\dot{\rho_{\phi}})$. In these forms,    $H$ is not directly involved, as it has the dimension of the inverse of time, and the latter   is already  present in $\dot\rho_i$.

In a spatially flat Friedmann-Lemaitre-Robertson-Walker (FLRW) Universe, the evolution equations are given by
\begin{eqnarray}
H^{2}&=&\frac{\kappa^2}{3} (\rho_{m}+\rho_{\phi})\nonumber \\
2\dot{H}+3H^2&=&-\kappa^2 p_{\phi}
\label{eq:H}
\end{eqnarray}
where $\kappa^2= 8\pi G$, $\rho_{\phi}=-\frac{1}{2}\dot{\phi^2}+ V(\phi)$ and 
$p_{\phi}=-\frac{1}{2}\dot{\phi^2}- V(\phi)$.
To cast the evolution equations into an autonomous system, we introduce the following dimensionless quantities, 
\begin{eqnarray}
x=\frac{\kappa\dot{\phi}}{\sqrt{6}H}; \quad y=\frac{\kappa \sqrt{V}}{\sqrt{3}H}; \quad
\lambda =-\frac{V'}{\kappa V}
\end{eqnarray}
Hence, we find 
\begin{eqnarray}
\dfrac{dx}{dN}&=& x \left( \frac{\ddot{\phi}}{H \dot{\phi}}-\frac{\dot{H}}{H^2}\right)  \nonumber\\
\dfrac{dy}{dN}&=&-y\left( \sqrt{\frac{3}{2}}\lambda x +\frac{\dot{H}}{H^{2}}\right) 
\label{eq:auto}
\end{eqnarray}
where, $N=\ln a$.  For an exponential potential,  we find that  $\lambda$ is constant, and
\begin{eqnarray}
\label{eq:hd}
\frac{\dot{H}}{H^2}&=& \frac{3(x^2+y^2-1)}{2}\\
\frac{\ddot{\phi}}{H \dot{\phi}}&=&-3-\sqrt{3/2}~\frac{\lambda y^2}{x}+\frac{Q}{H\dot{\phi^2}}
\label{eq:phidd}
\end{eqnarray}
Then, the effective equation of state, the field density parameter and the equation of state for a phantom field are given, respectively,  by
\begin{eqnarray}
\label{eq:weff}
w_{eff}&=& -1 - \frac{2\dot{H}}{3H^2} \nonumber\\
\Omega_{\phi}&=&\frac{\kappa^2 \rho_{\phi}}{3H^2}=-x^2+y^2 \nonumber\\
w_{\phi}&=&\frac{w_{eff}}{\Omega_{\phi}}
\end{eqnarray}
For an accelerating Universe, we have  $w_{eff} < -\frac{1}{3}$.

\section{Stationary points and their stabilities}
\label{sec:stat}

To study stationary points and their stabilities, let us consider the  autonomous system (\ref{eq:auto}), from which we can find the stationary points by setting the left-hand side of these equations to zero. Then, the signs of the eigenvalues will
tell us the stability of the points. In the following subsections, we consider different forms of the coupling.

\subsection{Coupling $Q=\alpha \dot{\rho}_{m}$}
\label{coup1}

For this  coupling, equation (\ref{eq:phidd}) takes the form,
\begin{eqnarray}
\frac{\ddot{\phi}}{H \dot{\phi}}&=&-3-\sqrt{3/2}~\frac{\lambda y^2}{x}-\frac{3\alpha \Omega_m}{2(1-\alpha)x^2}
\label{eq:phidd1}
\end{eqnarray}
where, $\Omega_m=1-\Omega_{\phi}$. Then, the autonomous system can be written as
\begin{eqnarray}
\dfrac{dx}{dN}&=& x \left( -3-\sqrt{3/2}~\frac{\lambda y^2}{x}-\frac{3\alpha \Omega_m}{2(1-\alpha)x^2}-\frac{3(x^2+y^2-1)}{2}\right) \nonumber \\
\dfrac{dy}{dN}&=&-y\left( \sqrt{\frac{3}{2}}\lambda x +\frac{3(x^2+y^2-1)}{2}\right)
\label{eq:auto1}
\end{eqnarray}
The critical points can be obtained by putting $\frac{dx}{dN}=0$ and $\frac{dy}{dN}=0$, simultaneously. Therefore, we have the following stationary points:

(1)~~$x= -\sqrt{\frac{\alpha}{\alpha-1}},~ y= 0$. In this case, the corresponding eigenvalues are,
\bqn
&& \mu_1 = -6 - \frac{3}{\alpha-1} \ < 0, ~~~~~~~~~~~~ \;\;\; {\mbox{ for}}\;\;\;  \alpha \ > 1, \nb\\
&&
\mu_2 = \frac{-3 + \sqrt{6 \alpha (\alpha-1)}~ \lambda}{2(\alpha-1)} \ < 0, \;\;\;  {\mbox{ for}} \;\;\; \alpha \ > 1,\; \sqrt{6 \alpha (\alpha-1)}~ \lambda \leq 0 \nb
\eqn
The point has negative eigenvalues for $\alpha \ > 1$ and $\sqrt{6 \alpha (\alpha-1)}~ \lambda \leq 0 $. Thus, it is a stable point.

(2)~~$x= \sqrt{\frac{\alpha}{\alpha-1}},~ y= 0$. Then, we have following eigenvalues,
\bqn
&& \mu_1 = -6 - \frac{3}{\alpha-1} \ < 0,   ~~~~~~~~~~~~ \;\;\; {\mbox{ for}}\;\;\;   \alpha \ > 1, \nb\\
&& \mu_2 = \frac{-3 - \sqrt{6 \alpha (\alpha-1)}~ \lambda}{2(\alpha-1)} \ < 0,   \;\;\; {\mbox{ for}}\;\;\;  \alpha \ > 1, \; \sqrt{6 \alpha (\alpha-1)}~ \lambda \geq 0 \nb
\eqn
The eigenvalues of this point show their negativity for $\alpha \ > 1$ and $\sqrt{6 \alpha (\alpha-1)}~ \lambda \geq 0 $. Therefore, it is a stable point.

(3)~~$x= -\frac{\lambda}{\sqrt{6}},~ y= -\sqrt{1+\frac{\lambda^{2}}{6}}$. In this case, the eigenvalues are given by,
\bqn
&&
\mu_1 = -3 - \lambda^2/2 \ < 0,  ~~~~~~~ \;\;\; {\mbox{ for}}\;\;\; \lambda \ > 0,\nb\\
&& \mu_2 = 3/(\alpha-1) - \lambda^2 \ < 0, ~~ \;\;\; {\mbox{ for}}\;\;\; \alpha \ > 1,\; \lambda \ > \sqrt{3/( \alpha-1)}\nb
\eqn
The point is stable under above given conditions.

(4)~~$x= -\frac{\lambda}{\sqrt{6}},~ y= \sqrt{1+\frac{\lambda^{2}}{6}}$. In this case, we get same eigenvalues as  (3).

(5)~~$x= \frac{\sqrt{\frac{3}{2}}}{\lambda(1-\alpha)},~ y= -\frac{\sqrt{(\alpha-1)\alpha \lambda^2-\frac{3}{2}}}{\lambda(\alpha-1)}$. In this case, the corresponding eigenvalues are,
\bqn
&& \mu_1 = -\frac{1}{4}(12+\frac{9}{\alpha-1}-2\alpha \lambda^2+\delta_1) < 0, \;\;\; {\mbox{ for}}\;\;\;  12+\frac{9}{\alpha-1}-2\alpha \lambda^2+\delta_1>0,\nb\\
&& \mu_2 = -\frac{1}{4}(12+\frac{9}{\alpha-1}-2\alpha \lambda^2-\delta_1) < 0, \;\;\; {\mbox{ for}}\;\;\;  12+\frac{9}{\alpha-1}-2\alpha \lambda^2-\delta_1>0,\nb
\eqn
where $\delta_1=\frac{\sqrt{(\alpha-1)\lambda^2(216+(\alpha-1)\lambda^2(-63+4\alpha(-54+36\alpha-3(\alpha-1)(4\alpha-5)\lambda^2+(\alpha-1)^2\alpha\lambda^4)))}}{\lambda^{2}(\alpha-1)^{2}}$.
The point now is a saddle point.

\begin{figure}[tbp]
{\includegraphics[width=2.3in,height=2.3in,angle=0]{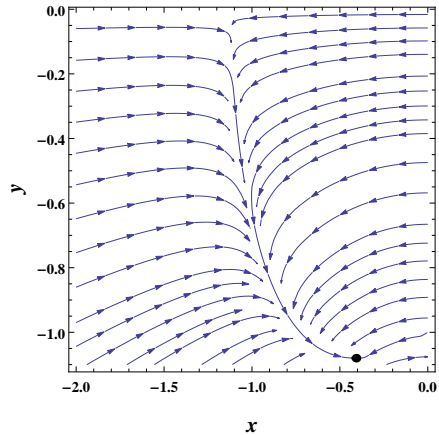}} 
\caption{The figure shows the phase space trajectories for point (3) of the coupling $Q=\alpha \dot{\rho}_{m}$. The stable fixed point is an attractive node and corresponds to $\alpha=5$ and $\lambda=1$. 
The black dot represents the stable attractor point. }
\label{figint1a}
\end{figure}
\begin{figure}[tbp]
\begin{center}
\begin{tabular}{ccc}
{\includegraphics[width=2.1in,height=2.1in,angle=0]{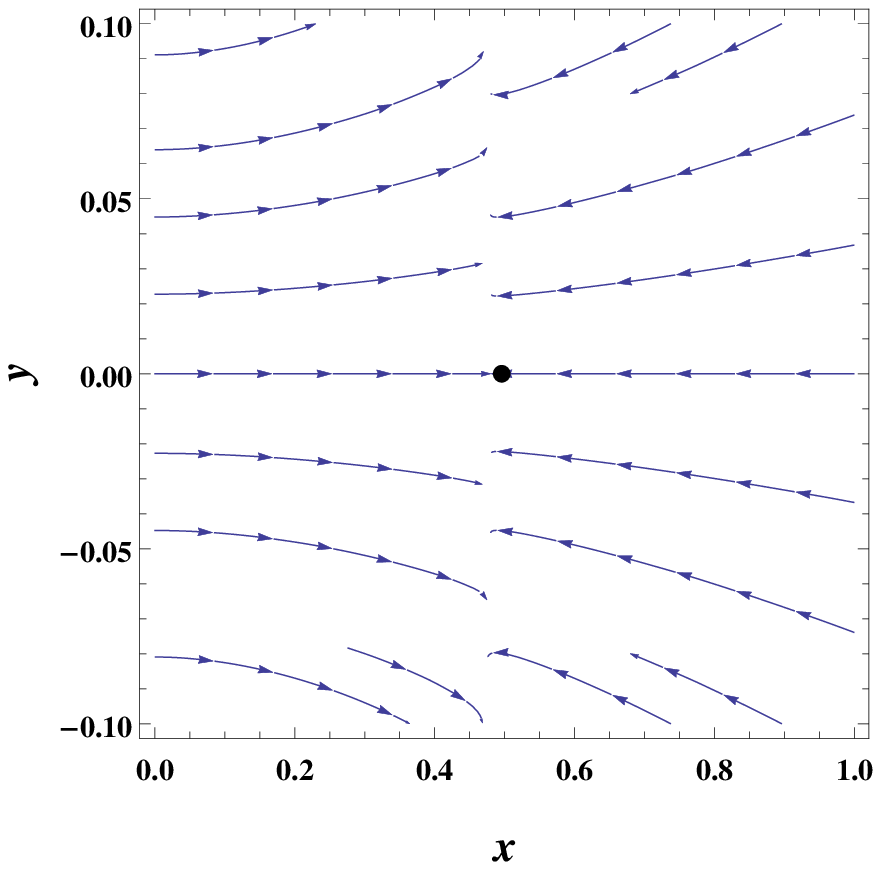}} & {%
\includegraphics[width=2.1in,height=2.1in,angle=0]{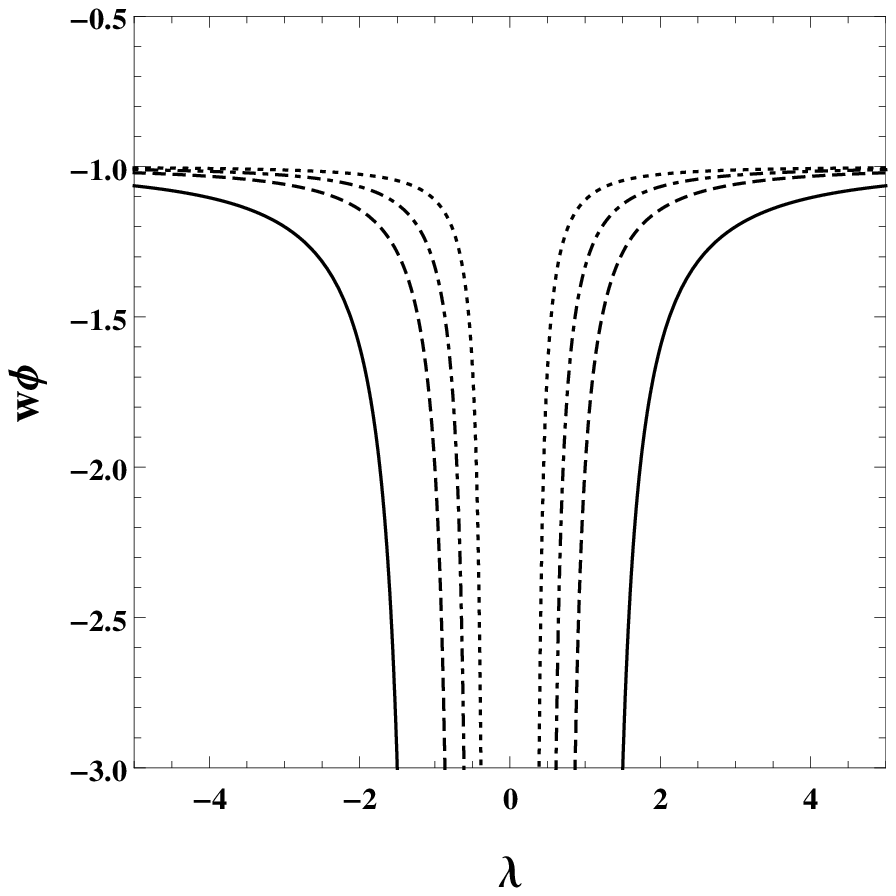}}  & {%
\includegraphics[width=2.1in,height=2.1in,angle=0]{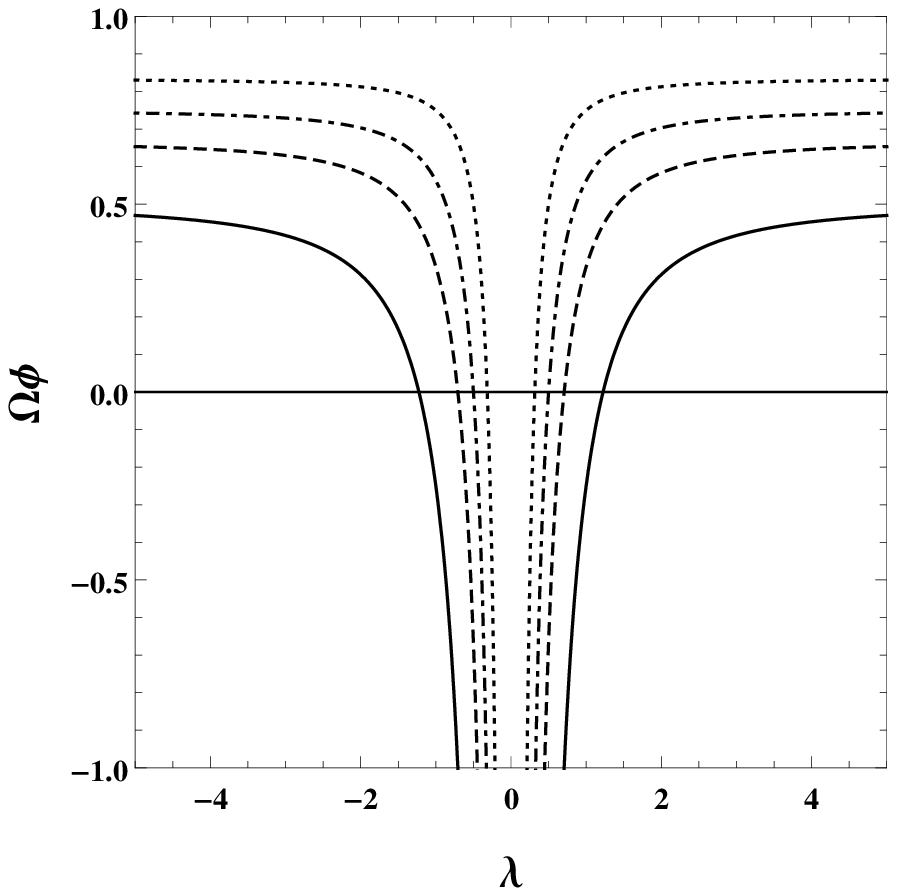}} 
\end{tabular}%
\end{center}
\caption{This figure represents the phase portrait, evolution of $w_{\phi}$ and $\Omega_{\phi}$ of point (5) for $Q=\alpha \dot{\rho}_{m}$. This is an unstable point and acts as a saddle point  that is shown in the left panel for $\alpha=-0.3$ and $\lambda=1.9$. The middle and right panels are plotted for different values of $\alpha$. The solid, dashed, dot-dashed and dotted lines correspond to $\alpha=-1, -2, -3$ and $-5$, respectively. The values of $\lambda$ below horizontal line are not allowed. }
\label{figint1b}
\end{figure}
\begin{table}
\caption{We display stationary points for the coupling $Q=\alpha\dot{\rho_m}$. We also show the expressions of $\Omega_{\phi}$, $w_{eff}$, $w_{\phi}$ and the conditions to have an  accelerating phase.}
\begin{center}
\label{tab1}
\begin{tabular}{l c c c c c c r} 
\hline\hline 
\\
Point &$x$  &$y$ &Stability &$\Omega_{\phi}$ &$w_{eff}$ & $w_{\phi}=\frac{w_{eff}}{\Omega_{\phi}}$ & Acceleration\\
\\
\hline
\\
1  &  $-\sqrt{\frac{\alpha}{\alpha-1}}$ &   $0$ & Stable for $\alpha>1$,  &  $\frac{\alpha}{1-\alpha}$  & $\frac{\alpha}{1-\alpha}$ & $1$ & No\\
&&&$\sqrt{6\alpha(\alpha-1)}\lambda \leq 0$&&&&\\
\\
\\
\hline
\\ 
2 & $\sqrt{\frac{\alpha}{\alpha-1}}$ &   $0$ &   Stable for $\alpha>1$,  &  $\frac{\alpha}{1-\alpha}$  & $\frac{\alpha}{1-\alpha}$ & $1$ & No\\
&&&$\sqrt{6\alpha(\alpha-1)}\lambda \geq 0$&&&&\\
\\
\\
\hline
\\ 
3,4 & $-\frac{\lambda}{\sqrt{6}}$ &   $\mp \sqrt{1+\frac{\lambda^2}{6}}$ &  Stable for $\alpha>1$,  &  $1$  & $-1-\frac{\lambda^2}{3}$ & $-1-\frac{\lambda^2}{3}$ & Yes\\
&&&$\lambda>\sqrt{\frac{3}{\alpha-1}}$&&&&\\
\\
\\
\hline
\\ 
5 & $\frac{\sqrt{3/2}}{\lambda(1-\alpha)}$ &   $-\frac{\sqrt{(\alpha-1)\alpha\lambda^2-3/2}}{\lambda(\alpha-1)}$ & Saddle for  &  $\frac{(\alpha-1)\alpha\lambda^2-3}{(\alpha-1)^2\lambda^2}$  & $\frac{\alpha}{1-\alpha}$ & $-\frac{(\alpha-1)\alpha\lambda^2}{(\alpha-1)\alpha\lambda^2-3}$ & Yes\\
&&&$12+\frac{9}{\alpha-1}-2\alpha\lambda^2>\delta_1$&&&&\\
\\
\\
\hline\hline 
\end{tabular}    
\end{center}
\end{table}

In this coupling, we are interested in Cases  (3) and (5),  as Case (3) is stable and has an accelerating period,  whereas Case (5) is a saddle point, and also has an accelerating period. For Case (3), we solve the autonomous system (\ref{eq:auto1}) numerically for $\alpha=5$ and $\lambda=1$, and the result is displayed in Fig. \ref{figint1a}. The stable point of Case (3) acts as an attractive node under the chosen parameters which is confirmed by Fig. \ref{figint1a}. Additionally, in this case we obtain $\Omega_{\phi}=1$ that corresponds to the case where dark energy totally dominates. However,  we find that  Case  (3) is a stable fixed point with a late accelerating Universe ($w_eff < -1/3$),  but it can not solve the coincidence problem as it has $\Omega_{DE}=1$ rather than $\Omega_{DE}/\Omega_{DM} \simeq {\cal{O}}(1)$. In Case (5), we evolve the system (\ref{eq:auto1}) numerically for $\alpha=-0.3$ and $\lambda=1.9$, and find that the nature of this point is a saddle point, which is shown at the left panel of Fig. \ref{figint1b}. We also find the cosmological observables $\Omega_{\phi}$, $w_{eff}$ and $w_{\phi}$. The middle and right panels of Fig. \ref{figint1b} show the evolution of $w_{\phi}$ and $\Omega_{\phi}$ versus $\lambda$. They also show for which range of $\lambda$ (having different values of $\alpha$) both physical observables are allowed. The general properties of this coupling are summarized in Table \ref{tab1}.

\subsection{Coupling $Q=\beta \dot{\rho_{\phi}}$}

\label{coup2}
For the coupling  $Q=\beta \dot{\rho_{\phi}}$, equation (\ref{eq:phidd}) becomes,
\begin{eqnarray}
\frac{\ddot{\phi}}{H \dot{\phi}}&=&-3-\sqrt{3/2}~\frac{\lambda y^2}{x}+\frac{3\beta }{1+\beta}
\label{eq:phidd2}
\end{eqnarray}
Therefore, equation (\ref{eq:auto}) takes the form,
\begin{eqnarray}
\dfrac{dx}{dN}&=& x \left( -3-\sqrt{3/2}~\frac{\lambda y^2}{x}+\frac{3\beta }{1+\beta}-\frac{3(x^2+y^2-1)}{2}\right)  \nonumber \\
\dfrac{dy}{dN}&=&-y\left( \sqrt{\frac{3}{2}}\lambda x +\frac{3(x^2+y^2-1)}{2}\right)
\label{eq:auto2}
\end{eqnarray}
For this coupling, we have the following stationary points:
\\
\\
\\
(1)~~$x= 0,~ y= 0$. In this case, the corresponding eigenvalues are,
\bqn
&& \mu_1 = \frac{3}{2}-\frac{3}{1+\beta} \ < 0, ~~~~~~~~~~~~ \;\;\; {\mbox{ for}}\;\;\;  0 \ < \beta \ < 1, \nb\\
&&
\mu_2 = \frac{3}{2} \nb
\eqn
As one of the eigenvalue is positive, the stationary point is unstable for any value of $\beta$.
\\
(2)~~$x= -\sqrt{\frac{\beta-1}{\beta+1}},~ y= 0$. In this case, the eigenvalues are given as,
\bqn
&& \mu_1 = -3+\frac{6}{1+\beta} \ < 0, ~~~~~~~~~~~~~~~~~~~~~~ \;\;\; {\mbox{ for}}\;\;\;  \beta \ < -1, \nb\\
&&
\mu_2 = \frac{3}{1+\beta}+\sqrt{\frac{3}{2}-\frac{3}{1+\beta}}\lambda \ < 0, ~~~~~~~~ \;\;\; {\mbox{ for}}\;\;\;  -2 \leq \beta \ < -1 \ \text{and} \ 0<\lambda \leq 1, \nb
\eqn
The eigenvalues of this point show the negativity for $-2 \leq \beta \ < -1$ and $0<\lambda \leq 1$. Therefore, it is a stable point.
\\
(3)~~$x= \sqrt{\frac{\beta-1}{\beta+1}},~ y= 0$. In this case, the corresponding eigenvalues are,
\bqn
&& \mu_1 = -3+\frac{6}{1+\beta} \ < 0, ~~~~~~~~~~~~~~~~~~~~~~ \;\;\; {\mbox{ for}}\;\;\;  \beta \ < -1, \nb\\
&&
\mu_2 = \frac{3}{1+\beta}-\sqrt{\frac{3}{2}-\frac{3}{1+\beta}}\lambda \ < 0, ~~~~~~~~ \;\;\; {\mbox{ for}}\;\;\;   \beta \ < -1 \ \text{and} \ \lambda > \ 0, \nb
\eqn
It is stable point for above given conditions.
\\
\\
(4)~~$x= \frac{9-(1+\beta)^2\lambda^4+\delta_2}{2\sqrt{6}\lambda(1+\beta)(3+(1+\beta)\lambda^2)},~ y= -\frac{\sqrt{6(1+\beta)^2\lambda^2-9+\lambda^4(1+\beta)^2-\delta_2}}{2\sqrt{3}(1+\beta)\lambda}$.
In this case,  we have following eigenvalues,
\bqn
&& \mu_1 = -\frac{2\delta_{2}^2-6(1+\beta)\delta_2 \epsilon \lambda^2+2\epsilon^2(-9+(1+\beta)\lambda^2(9+2(1+\beta)\lambda^2))+\nu}{16(1+\beta)^2\epsilon^2\lambda^2} \ < 0, \nb \\ &&~~~~ \;\;\; {\mbox{ for}}\;\;\;  2\delta_{2}^2+2\epsilon^2(-9+(1+\beta)\lambda^2(9+2(1+\beta)\lambda^2))+\nu \ < 0, \nb\\ \nb \\ \nb \\
&&
\mu_2 = -\frac{2\delta_{2}^2-6(1+\beta)\delta_2 \epsilon \lambda^2+2\epsilon^2(-9+(1+\beta)\lambda^2(9+2(1+\beta)\lambda^2))-\nu}{16(1+\beta)^2\epsilon^2\lambda^2} \ < 0,\nb \\ &&~~~~ \;\;\; {\mbox{ for}}\;\;\;   2\delta_{2}^2+2\epsilon^2(-9+(1+\beta)\lambda^2(9+2(1+\beta)\lambda^2))-\nu \ < 0, \nb
\eqn
where,
\bqn
&& \delta_2=\sqrt{(3+(1+\beta)\lambda^2)^2(9+(1+\beta)\lambda^2(6+\lambda^2+\beta(12+\lambda^2)))}\nb
\\ 
&& \epsilon=3+(1+\beta)\lambda^2 \nb
\\
&& \nu=\surd \left( \delta_{2}^4-12(1+\beta)\lambda^2\delta_{2}^3\epsilon-6\delta_{2}^2\epsilon^2(3-4(\beta-6)(1+\beta)\lambda^2+3(1+\beta)^2\lambda^4)+\epsilon^4(-9+12(1+\beta)(2+\beta)\lambda^2
\right. \nb \\  && \left.
~~~~ +5(1+\beta)^2\lambda^4)^2+4(1+\beta)\lambda^2\delta_2\epsilon^3(-117+(1+\beta)\lambda^2(24+\lambda^2+\beta(60+\lambda^2))))\right)\nb
\eqn
The eigenvalues of this point show the negativity under above conditions. Hence, it is a stable point. The $\Omega_{\phi}$, $w_{eff}$ and $w_{\phi}$ are given as,
\begin{eqnarray}
\label{eq:2omega}
\Omega_{\phi}&=&-\frac{9-(1+\beta)^2 \lambda^4+\sqrt{\delta_2}}{2(1+\beta)^2 \lambda^2 \epsilon} \\
\label{eq:2weff}
w_{eff}&=&-\frac{9+18\beta+6(1+\beta)^2 \lambda^2+(1+\beta)^2 \lambda^4-\sqrt{\delta_2}}{6(1+\beta) \epsilon} \\
w_{\phi}&=&\frac{-9-(1+\beta) \lambda^2(6+(1+\beta) \lambda^2)+\sqrt{\gamma}}{6 \epsilon} 
\label{eq:2w}
\end{eqnarray} 
where,
\begin{eqnarray}
\gamma &=& 81+(1+\beta)^2 \lambda^2 \Big{(}108+18(3+4\beta)\lambda^2+12(1+\beta)^2 \lambda^4+(1+\beta)^2 \lambda^6 \Big{)} 
\end{eqnarray}
\begin{figure}[tbp]
{\includegraphics[width=2.3in,height=2.3in,angle=0]{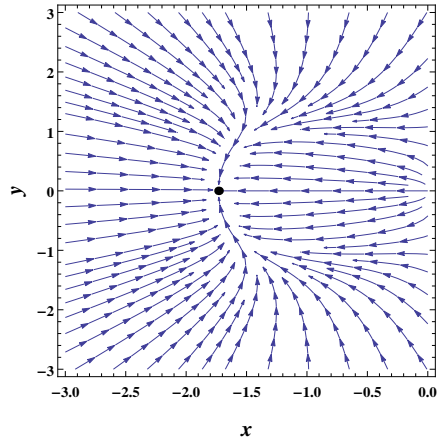}}
\caption{The figure displays the phase space trajectories of Case (2) for $Q=\beta \dot{\rho_{\phi}}$. It is plotted for $\beta=-2$ and $\lambda=1$. The point is stable and behaves as an attractive node.}
\label{figint2a}
\end{figure}
\begin{figure}[tbp]
\begin{center}
\begin{tabular}{ccc}
{\includegraphics[width=2.1in,height=2.1in,angle=0]{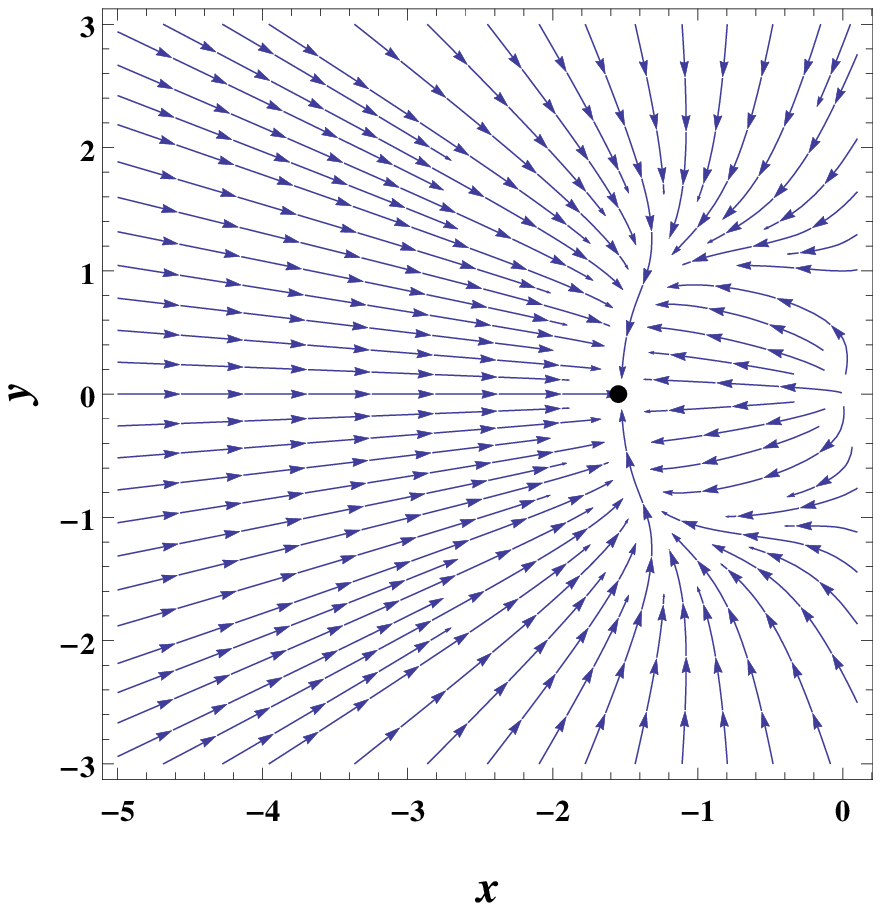}} & {%
\includegraphics[width=2.1in,height=2.1in,angle=0]{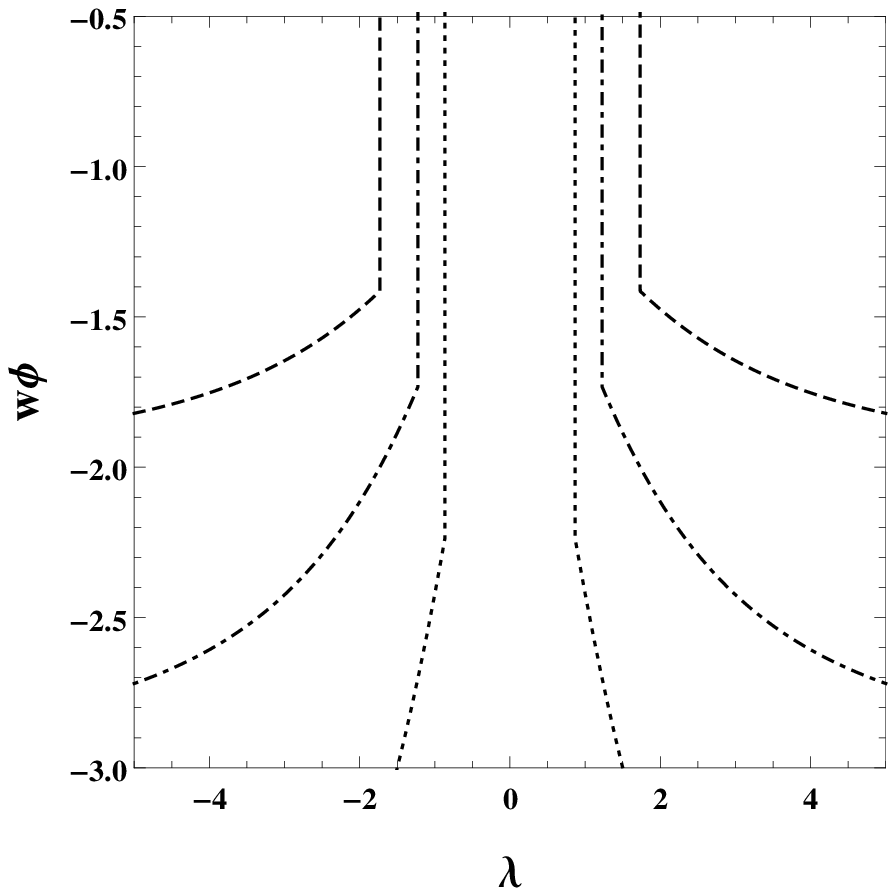}}  & {%
\includegraphics[width=2.1in,height=2.1in,angle=0]{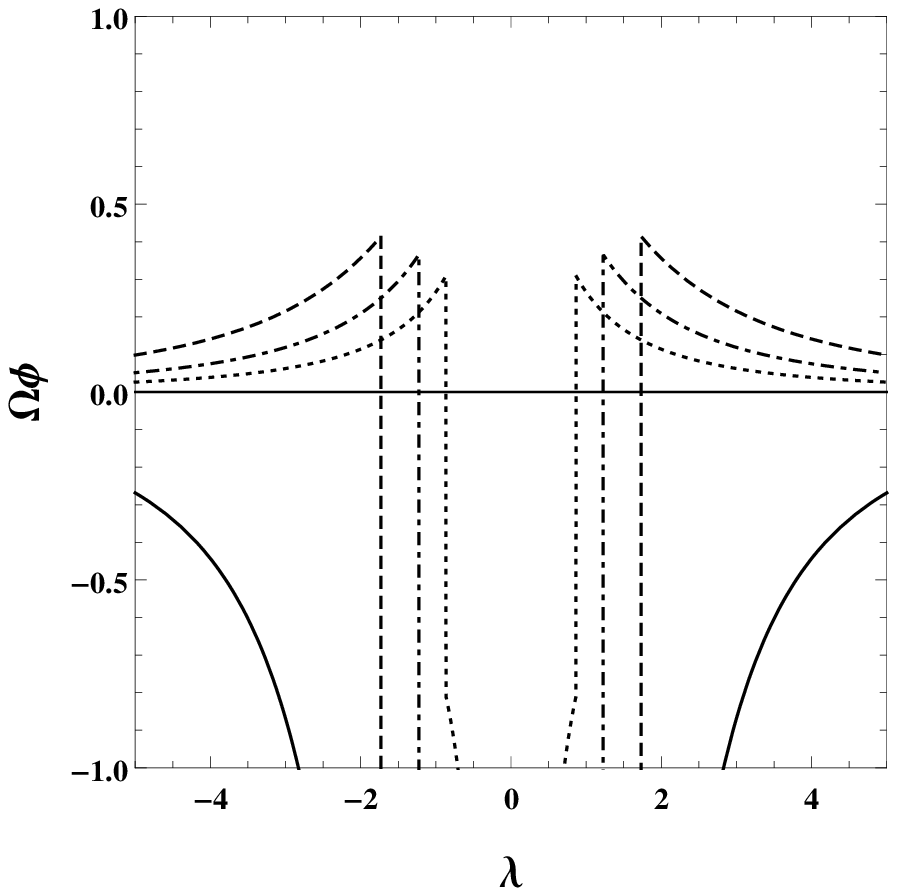}} 
\end{tabular}%
\end{center}
\caption{The left panel shows the phase portrait of Case (4) for $Q=\beta \dot{\rho_{\phi}}$, and corresponds to $\beta=-2.5$ and $\lambda=1$. The middle and right panels show the evolution of 
$w_{\phi}$ and $\Omega_{\phi}$ versus $\lambda$ for various values of $\beta$. The solid, dashed, dot-dashed and dotted lines correspond to $\beta=-0.5, -2, -3$ and $-5$, respectively. The  values of $\lambda$ below the horizontal line are not accepted. This is a stable point and acts as an attractive node. }
\label{figint2b}
\end{figure}

\begin{table}
\caption{We present stationary points and their stability for the coupling $Q=\beta\dot{\rho_{\phi}}$.}
\begin{center}
\label{tab2}
\begin{tabular}{l c c c c c c r} 
\hline\hline 
\\
Point &$x$  &$y$ &Stability &$\Omega_{\phi}$ &$w_{eff}$ & $w_{\phi}=\frac{w_{eff}}{\Omega_{\phi}}$ & Accele-\\
&&&&&&& ration
\\
\\
\hline
\\
1  &  $0$ &   $0$ & Saddle  &  $0$  & $0$ & Indeterminate & No\\
\\
\\
\hline
\\ 
2& $- \sqrt{\frac{\beta-1}{\beta+1}}$ &   $0$ &   Stable for $-2 \leq \beta<-1$  &  $\frac{1-\beta}{1+\beta}$  & $\frac{1-\beta}{1+\beta}$ & $1$ & No\\
&&& and $0< \lambda \leq 1$
\\
\\
\hline
\\ 
3 & $ \sqrt{\frac{\beta-1}{\beta+1}}$ &   $0$ &   Stable for $\beta<-1$  &  $\frac{1-\beta}{1+\beta}$  & $\frac{1-\beta}{1+\beta}$ & $1$ & No\\
&&& and $\lambda > 0$
\\
\\
\hline
\\ 
4 & $\frac{9-(1+\beta)^2\lambda^4+\delta_2}{2\sqrt{6}\lambda(1+\beta)(3+(1+\beta)\lambda^2)}$ &   $-\frac{\sqrt{6(1+\beta)^2\lambda^2-9+\lambda^4(1+\beta)^2-\delta_2}}{2\sqrt{3}(1+\beta)\lambda}$ & $-$  &  Eq.(\ref{eq:2omega})  & Eq.(\ref{eq:2weff}) & Eq.(\ref{eq:2w}) & $-$\\
\\
&&&(a) Stable for $\beta=-2.5$ &&& Positive & No\\
&&& and $1 \leq \lambda < 1.5$ \\
\\
&&&(b) Saddle for $\beta=-2.5$ &&& $<-1$ & Yes\\
&&& and $\lambda \geq 1.5$ \\
\\
\\
\hline\hline 
\end{tabular}    
\end{center}
\end{table}
For this coupling, we pay particular attention on Cases (2) and (4). In case (2), we evolve the autonomous system (\ref{eq:auto2}) numerically for the values $\beta=-2$ and $\lambda=1$, and get $\Omega_{\phi}$, $w_{eff}$ and $w_{\phi}$. With the chosen parameters, the point is stable and behaves as an attractive node (see Fig. \ref{figint2a}), but there does not exist an  accelerating phase  of the Universe, as the equation of state $w_{\phi}$ for phantom field is always positive. Therefore, it does not solve the coincidence problem. In Case (4), we elaborate the system for $\beta=-2.5$ and $\lambda=1$, and find that it is stable and acts as an attractive node. The phase portrait of this stable point is shown in the left panel of Fig. \ref{figint2b}, the middle and right panels of Fig. \ref{figint2b} show the evolution of $w_{\phi}$ and $\Omega_{\phi}$. For this point, we consider two cases: (a) $\beta=-2.5$ and $1 \leq \lambda < 1.5$, in which case (4) behaves as a stable point but does not give rise to an accelerating Universe as $w_{\phi}$ is always positive. (b) $\beta=-2.5$ and $\lambda > 1.5$, in which Case (4) acts as a saddle point and has an accelerating phase as $w_{\phi}<-1$ (see Table \ref{tab2}). Hence,  it does not alleviate  the coincidence problem. The results of the coupling are summarized  in Table \ref{tab2}.

\subsection{Coupling $Q=\sigma (\dot{\rho}_{m}+ \dot{\rho_{\phi}})$}
\label{coup3}

In this case, the coupling $Q$ is a linear combination of $\dot{\rho}_{m}$ and $\dot{\rho_{\phi}}$. For this coupling, equation (\ref{eq:phidd}) can be written as,
\begin{eqnarray}
\frac{\ddot{\phi}}{H \dot{\phi}}&=&-3-\sqrt{3/2}~\frac{\lambda y^2}{x}-\frac{3\sigma \Omega_m }{2(1-\sigma)x^2}+\frac{3\sigma}{1+\sigma},
\label{eq:phidd3}
\end{eqnarray}
Thus, the autonomous system (\ref{eq:auto}) becomes,
\begin{eqnarray}
\dfrac{dx}{dN}&=& x \left( -3-\sqrt{3/2}~\frac{\lambda y^2}{x}-\frac{3\sigma \Omega_m }{2(1-\sigma)x^2}+\frac{3\sigma}{1+\sigma}-\frac{3(x^2+y^2-1)}{2}\right) \nonumber \\
\dfrac{dy}{dN}&=&-y\left( \sqrt{\frac{3}{2}}\lambda x +\frac{3(x^2+y^2-1)}{2}\right)
\label{eq:auto3}
\end{eqnarray}
For this coupling, we have the following stationary points:
\\
\\
(1)~~$x= -\frac{\sqrt{1-\sigma+2\sigma^2-\sqrt{1+\sigma(\sigma+8\sigma^3-6)}}}{\sqrt{2(\sigma^2-1)}},~ y= 0$. In this case, the corresponding eigenvalues are, 
\bqn
&& \mu_1 = \frac{3\sqrt{1+\sigma(\sigma+8\sigma^3-6)}}{\sigma^2-1}<0, ~~ \text{for} \ \sigma^2 \ < 1, \nb
\\
&& \mu_2 = \frac{1}{8}\left( \frac{6(\sigma-3+\sqrt{1+\sigma(\sigma+8\sigma^3-6)})}{\sigma^2-1}+4\sqrt{3}\lambda\sqrt{\frac{1-\sigma+2\sigma^2-\sqrt{1+\sigma(\sigma+8\sigma^3-6)}}{\sigma^2-1}}\right)>0,~ \text{for all}\ \sigma. \nb
\eqn
As one of the eigenvalue is positive, the stationary point is a saddle for any value of $\sigma$.
\\
\\
(2)~~$x= \frac{\sqrt{1-\sigma+2\sigma^2-\sqrt{1+\sigma(\sigma+8\sigma^3-6)}}}{\sqrt{2(\sigma^2-1)}},~ y= 0$. In this case, the eigenvalues are given as, 
\bqn
&& \mu_1 = \frac{3\sqrt{1+\sigma(\sigma+8\sigma^3-6)}}{\sigma^2-1}<0,~~ \text{for}\ \sigma^2 \ < 1, \nb
\\
&& \mu_2 = \frac{1}{8}\left( \frac{6(\sigma-3+\sqrt{1+\sigma(\sigma+8\sigma^3-6)})}{\sigma^2-1}-4\sqrt{3}\lambda\sqrt{\frac{1-\sigma+2\sigma^2-\sqrt{1+\sigma(\sigma+8\sigma^3-6)}}{\sigma^2-1}}\right)>0,~ \text{for all} \ \sigma.\nb
\eqn
This is a saddle point.
\\
\\
(3)~~$x= \sqrt{\frac{1+\sigma(2\sigma-1)+\sqrt{1+\sigma(\sigma+8\sigma^3-6)}}{2(\sigma^2-1)}},~ y= 0$. In this case, the eigenvalues take the form, 
\bqn
&& \mu_1 = -\frac{3\sqrt{1+\sigma(\sigma+8\sigma^3-6)}}{\sigma^2-1}<0,~~ \text{for}  
\ \sigma^2 \ > 1, \nb
\\
&& \mu_2 = \frac{1}{8}\left( -\frac{6(3-\sigma+\sqrt{1+\sigma(\sigma+8\sigma^3-6)})}{\sigma^2-1}-4\sqrt{3}\lambda\sqrt{\frac{1+\sigma(2\sigma-1)+\sqrt{1+\sigma(\sigma+8\sigma^3-6)}}{\sigma^2-1}}\right)<0,\nb \\
&&~~~~~~~ \text{for}\ \sigma^2 > 1 \ \text{and}\ \lambda \ >0.\nb
\eqn
The eigenvalues of the point show the negativity for $\sigma^2 > 1$ and $\lambda>0$. Therefore, it is a stable point.

\begin{figure}[tbp]
{\includegraphics[width=2.3in,height=2.3in,angle=0]{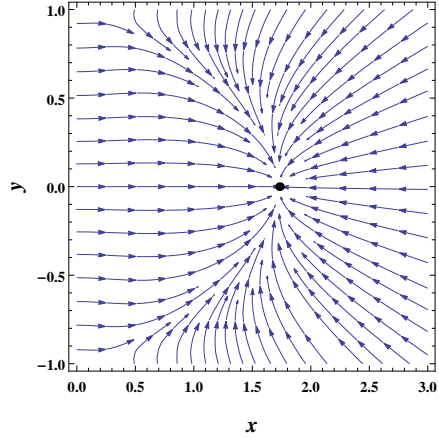}} 
\caption{The figure represents the evolution of the phase space trajectories of Case (3) for $Q=\sigma (\dot{\rho}_{m}+ \dot{\rho_{\phi}})$, and is plotted for $\sigma=2$ and $ \lambda=1$. The stable point acts as an attractive node, and the black dot designates  a stable attractor point.}
\label{figint3}
\end{figure}

For this  coupling, the stationary point in Case (3) is stable for $\sigma^2>1$ and $\lambda>0$. We numerically evolve the autonomous system (\ref{eq:auto3}) for the choices $\sigma =2$ and $\lambda=1$. The phase space trajectories of the stable point is displayed in Fig. \ref{figint3}, and the point behaves as an attractive node. For this point we do not find any accelerating solution as it has positive equation of state. Hence, it can not solve the coincidence problem. The main results of this coupling are summarized in Table \ref{tab3}.

In ref. \cite{alamepjc}, we studied the coupled quintessence with scaling potential for different forms of the coupling and discussed phase space analysis. For all the models, we obtained late time accelerated scaling attractor having $\Omega_{DE}/\Omega_{DM}= O(1)$. Therefore all the models considered in the said reference are viable to solve the coincidence problem. In the present paper, we perform same analysis with the coupled phantom field and inspect whether the coincidence problem can be alleviated or not. In case of coupling term $Q=\alpha \dot{\rho}_m$, the point (3) is a stable fixed point with an accelerating phase, but it can not solve the coincidence problem as $\Omega_{DE}=1$ (see Table \ref{tab1}). In case of $Q=\beta \dot{\rho}_{\phi}$, we focus on points (2) and (4), and notice that both are unable to solve the coincidence problem (see Table \ref{tab2}). In case of $Q=\sigma (\dot{\rho}_m + \dot{\rho}_{\phi})$, the point (3) is stable with non-accelerating phase as equation of state is positive (see Table \ref{tab3}). Therefore, in the interacting phantom field models, coincidence problem can not be solved. Similar results were discussed in ref. \cite{saridakisa}.
\begin{table}
\caption{We show stationary points for the coupling  $Q=\sigma (\dot{\rho}_{m}+ \dot{\rho_{\phi}})$.}
\begin{center}
\label{tab3}
\begin{tabular}{l c c c c c c r} 
\hline\hline 
\\
Point &$x$  &$y$ &Stability &$\Omega_{\phi}$ &$w_{eff}$ & $w_{\phi}=\frac{w_{eff}}{\Omega_{\phi}}$ & Acceleration\\
\\
\hline
\\
1, 2  &  $\mp \frac{\sqrt{1-\sigma+2\sigma^2-\sqrt{1+\sigma(\sigma+8\sigma^3-6)}}}{\sqrt{2}(\sigma^2-1)}$ &   $0$ & Saddle  &  $\frac{\sigma-1-2\sigma^2}{2(\sigma^2-1)}$  & $\frac{\sigma-1-2\sigma^2}{2(\sigma^2-1)}$ & 1 & No\\
&&&& $\frac{+\sqrt{1+\sigma(\sigma+8\sigma^3-6)}}{2(\sigma^2-1)}$ & $\frac{+\sqrt{1+\sigma(\sigma+8\sigma^3-6)}}{2(\sigma^2-1)}$
\\
\\
\hline
\\ 
3 & $\sqrt{\frac{1+\sigma(2\sigma-1)+\sqrt{1+\sigma(\sigma+8\sigma^3-6)}}{2(\sigma^2-1)}}$ &   $0$ &   Stable for   &  $\frac{1+\sigma(2\sigma-1)}{2(1-\sigma^2)}$   & $\frac{1+\sigma(2\sigma-1)}{2(1-\sigma^2)}$  & $1$ & No\\
&&&$\sigma^2>1$, $\lambda>0$ & $\frac{+\sqrt{1+\sigma(\sigma+8\sigma^3-6)}}{2(1-\sigma^2)}$ & $\frac{+\sqrt{1+\sigma(\sigma+8\sigma^3-6)}}{2(1-\sigma^2)}$
\\
\\
\\
\hline\hline 
\end{tabular}    
\end{center}
\end{table}
\section{Coupled tachyon dynamics}
\label{sec:ctd}
Tachyon acts as a source of dark energy, depending on the shape of the potentials \cite{paddy}.  We consider that dark energy and dark matter are interacting to each other, but the total energy density is conserved.  The conservation equations for both  components are written as,
\begin{eqnarray} 
\dot{\rho}_{{m}}+3H(\rho_{m}+p_m)=Q, \nonumber \\
\dot{\rho}_{{\phi}}+3H(\rho_{\phi}+p_{\phi})=-Q,   
\label{tconser}
\end{eqnarray}
where, 
\begin{equation}
\label{tep}
\rho_{\phi}=\frac{V(\phi)}{\sqrt{1-\dot{\phi}^{2}}}, \quad p_{\phi}=-V(\phi)\sqrt{1-\dot{\phi}^{2}}
\end{equation}
Then, the evolution equations take the form,
\begin{equation}\label{tH}
  H^{2}=\frac{\kappa^{2}}{3}\left[\frac{V(\phi)}{\sqrt{1-\dot{\phi}^{2}}} + \rho_{m} \right],
\end{equation}
\begin{equation}\label{tphidd}
  \frac{\ddot{\phi}}{1-\dot{\phi}^2}+3H\dot{\phi}+\frac{V'(\phi)}{V(\phi)}=-\frac{Q\sqrt{1-\dot{\phi}^2}}{\dot{\phi}V(\phi)}
\end{equation}
where a prime and a dot denote derivative with respect to field and cosmic time,  respectively.

Let us define the following dimensionless parameters
\begin{equation}\label{t5}
  x=\dot{\phi}, \quad y=\frac{\kappa\sqrt{V}}{\sqrt{3}H}, \quad \Omega_m=\frac{\kappa^2 \rho_{m}}{3H^2}, \quad \lambda=-\frac{V'}{\kappa V\sqrt{V}}
\end{equation}
Then, we obtain the autonomous system, 
\begin{eqnarray}
\frac{dx}{dN}&=& \frac{\ddot{\phi}}{H\dot{\phi}}~x \nonumber\\
\frac{dy}{dN}&=& -\frac{\sqrt{3}}{2}y^2\lambda x -y\left(\frac{\dot{H}}{H^2}\right)
\label{tauto}
\end{eqnarray}
Here we take inverse square potential for which $\lambda$ is constant. Also, we consider the coupling  $Q=\beta \dot{\rho_{\phi}}$ only. For this coupling we have following equations,
\begin{align}
\frac{\dot H}{H^2}&=\frac{3\left( y^2\sqrt{1-x^2}-1\right) }{2},\\
\nonumber\\
\frac{\ddot{\phi}}{H\dot{\phi}}&=-3\left( 1-x^2 \right) +\sqrt{3}\lambda y\frac{\left( 1-x^2\right) }{x}+\frac{3 \beta \left( 1-x^2\right)}{1+\beta},
\end{align}
The equation of state for the tachyon field is given as,
\begin{align}
w_{eff}&= -1 -\frac{2\dot H}{3H^2},\\
w_{\phi}&= \frac{w_{eff}-w_m \Omega_m}{1-\Omega_m},
\end{align}
\begin{figure}[tbp]
{\includegraphics[width=2.3in,height=2.3in,angle=0]{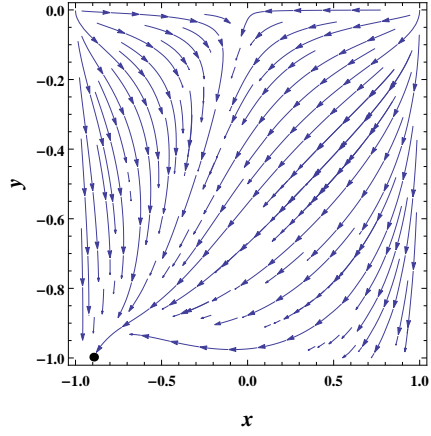}} 
\caption{The figure shows the phase space trajectories for the coupled tachyon field. The stable fixed point is an attractive node and corresponds to $\beta=0.9$ and $\lambda=1$. 
The black dot represents the stable attractor point. }
\label{figtech}
\end{figure}
where $w_m = 0$ for standard dust matter.
Setting  the left hand sides of the autonomous system (\ref{tauto}) to zero, we obtain the following stationary points:
\\
\\
(1)~~$x= 0,~ y= 0$. In this case, the corresponding eigenvalues are, 
\bqn
&& \mu_1 = -\frac{3}{1+\beta}<0,~~ \text{for} \ 0 < \ \beta \ < 1, \nb\\
&& \mu_2 = \frac{3}{2},\nb
\eqn
As one of the eigenvalue is positive, the stationary point is a saddle.
\\
\\
(2)~~$x= \pm 1,~ y= \pm \frac{\sqrt{3}}{\lambda}$. In this case, the metric is in-determinant. \\
\\
\\
(3)~~$x=-\dfrac{1}{3\sqrt{6}} (1 + \beta) \lambda  \surd \left( ((-81(2\times2^{1/3} \delta^{2/3}_{3} + 18(18+ \delta_{5})) +2^{2/3}\delta_{3}^{1/3}(18+\delta_{5}))+ 9(1+\beta)^{2} (243\beta^{2}(18+2^{2/3}\delta^{1/3}_{4})\right. \\ \left. \qquad \qquad
+  12\beta(486+27\times 2^{2/3}\delta_{4}^{1/3}-2^{1/3}\delta_{4}^{2/3}) + 135\times 2^{2/3}\delta_{4}^{1/3} -8\times 2^{1/3}\delta_{4}^{2/3} +18(153+\delta_{5}))\lambda^{4}
\right. \\ \left. \qquad \qquad
- 2(1+\beta)^{4}(243+2187\beta^{2}-81\beta(-18+2^{2/3}\delta_{4}^{1/3})-54 \times 2^{2/3}\delta_{4}^{1/3}+2^{1/3}\delta_{4}^{2/3})\lambda^{8} +2(1+\beta)^{6}
\right. \\ \left. \qquad \qquad
(-90-162\beta+2^{2/3}\delta_{4}^{1/3})\lambda^{12}-4(1+\beta)^{8}\lambda^{16} 
 \right. \\ \left. \qquad \qquad
  \Big {/}  ((1+\beta)^{2}\lambda^{2}(-81(18+\delta_{5}) +243(1+\beta)^{2}
 (5+3\beta(4+3\beta))\lambda^{4}+54(1+\beta)^{4}(2+3\beta)\lambda^{8}+2(1+\beta)^{6}\lambda^{12}))) \right) $\\
\\
\\
$y =  -\dfrac{1}{3\sqrt{2}}  \surd \left( ((-81(324+18\times 2^{2/3}\delta_{3}^{1/3} +2\times 2^{1/3}\delta_{3}^{2/3} +18\delta_{5} +2^{2/3}\delta_{3}^{1/3}\delta_{5}) +  
9(1+\beta)^{2}(2754+243\beta^{2}
\right. \\ \left. \qquad \qquad
(18+2^{2/3}\delta_{4}^{1/3})+12\beta(486 +27\times 2^{2/3}\delta_{4}^{1/3}-2^{1/3}\delta_{4}^{2/3})+135\times 2^{2/3}\delta_{4}^{1/3}- 8\times 2^{1/3}\delta_{4}^{2/3} +18\delta_{5})\lambda^{4}-2(1+\beta)^{4}
\right. \\ \left. \qquad \qquad
(243+2187\beta^{2}-81\beta(-18+2^{2/3}\delta_{4}^{1/3})-54\times 2^{2/3}\delta_{4}^{1/3} +2^{1/3}\delta_{4}^{2/3})\lambda^{8}
\right. \\ \left. \qquad \qquad
 +2(1+\beta)^{6}(-90-162\beta+2^{2/3}\delta_{4}^{1/3})
\lambda^{12}-4(1+\beta)^{8}\lambda^{16}) 
\right. \\ \left. \qquad \qquad
  \Big {/} ((1+\beta)^{2}\lambda^{2}(-81(18+\delta_{5})+243(1+\beta)^{2}(5+3\beta(4+3\beta))\lambda^{4}+54(1+\beta)^{4}(2+3\beta)\lambda^{8}+2(1+\beta)^{6}\lambda^{12}))) \right)   $
\\
\\
For this point, we get following eigenvalues,
\\
\\
\\
$\mu_1 = \dfrac{-\dfrac{\eta_{2}}{2}+3\beta \left(\dfrac{-\eta_{2}}{6}+\sqrt{\eta_{6}}\right)+\dfrac{\eta_{4} \left(\dfrac{\eta_{3}}{18}+2\sqrt{\eta 6}\right) }{6\eta 5} -3\sqrt{\eta_{6}}-\eta_{8}-\dfrac{1}{3}(1+\beta)^{2}\sqrt{\eta_{2}}\sqrt{\eta_{6}}\sqrt{\eta_{7}}\lambda^{2} }{4(1+\beta)\sqrt{\eta_{6}}}<0$\\
\\

~~~~~for $\beta \neq -1$ and $ 3\beta \left(\dfrac{-\eta_{2}}{6}+\sqrt{\eta_{6}}\right)+\dfrac{\eta_{4} \left(\dfrac{\eta_{3}}{18}+2\sqrt{\eta 6}\right) }{6\eta 5} <0 $
\\
\\
\\
\\
$\mu_{2}=\dfrac{\dfrac{\eta_{10}}{2}+3\beta \left(\dfrac{\eta_{10}}{6}+\sqrt{\eta_{6}}\right)+\dfrac{\eta_{4} \left(\dfrac{\eta_{3}}{18}+2\sqrt{\eta 6}\right) }{6\eta_{5}} -3\sqrt{\eta_{6}}+\eta_{8}-      \dfrac{(1+\beta)^{2}\sqrt{\eta2} \sqrt{3- \dfrac{\eta_{4}}{18\eta_{5}}}\sqrt{\eta_{7}}\lambda^{2}}{3\sqrt{3}} }{4(1+\beta)\sqrt{\eta_{6}}}<0$
\\
\\

~~~~~for $\beta \neq -1$ and $ \dfrac{\eta_{10}}{2}+3\beta \left(\dfrac{\eta_{10}}{6}+\sqrt{\eta_{6}}\right)+\dfrac{\eta_{4} \left(\dfrac{\eta_{3}}{18}+2\sqrt{\eta 6}\right) }{6\eta 5} +\eta_8<0 $
\\
\\
where,
\\
\\
$\delta_3=-243 (1+\beta )^2 \left (5+3 \beta  (4+3 \beta )\right) \lambda ^4-54 (1+\beta )^4 (2+3 \beta ) \lambda ^8-2 (1+\beta )^6 \lambda ^{12}+81 (18+\delta_5)$
\\
\\
$\delta_4=1458-243 (1+\beta )^2 (5+3 \beta  (4+3 \beta )) \lambda ^4-54 (1+\beta )^4 (2+3 \beta ) \lambda ^8-2 (1+\beta )^6 \lambda ^{12}+81 \delta_5$
\\
\\
$\delta_5=\sqrt{3 (1+\beta )^4 \lambda ^4 \left(-324+9 (1+\beta )^2 (-1+9 \beta  (2+3 \beta )) \lambda ^4+4 \beta  (1+\beta )^4 \lambda^8 \right)}$
\\
\\
$\eta_{1}= (-81(324 +18 \times 2^{2/3}\delta_{3}^{1/3} + 2\times 2^{1/3} \delta_{3}^{2/3} +18\delta_{5} +2^{2/3}\delta_{3}^{1/3} \delta_{5})+ 9(1+\beta)^{2}(2754+243\beta^{2}(18+2^{2/3}\delta_{4}^{1/3})
\\
\\
~~~~~~~
+12\beta(486+27\times 2^{2/3}\delta_{4}^{1/3}-2^{1/3}\delta_{4}^{2/3})+ 135\times 2^{2/3}\delta_{4}^{1/3}- 8\times 2^{1/3}\delta_{4}^{2/3}+18\delta_{5})\lambda^{4}-2(1+\beta)^{4}
\\
\\ 
~~~~~~~
(243+2187\beta^{2}-81\beta(-18+2^{2/3}\delta_{4}^{1/3})-54\times2^{2/3}\delta_{4}^{1/3}+2^{1/3}\delta_{4}^{2/3})\lambda^{8}+ 2(1+\beta)^{6}(-90-162\beta +2^{2/3}\delta_{4}^{1/3})\lambda^{12}
\\
\\ 
~~~~~~~
-4(1+\beta)^{8}\lambda^{16}$
\\
\\
$\eta_{2}=   \dfrac{\eta_{1}}{(1+\beta)^{2}\lambda^{2}(-81(18+\delta_{5})+243(1+\beta)^{2}(5+3\beta(4+3\beta))\lambda^{4}                                                                +54(1+\beta)^{4}(2+3\beta)\lambda^{8}+2(1+\beta)^{6}\lambda^{12})    }$
\\
\\
\\
$\eta_{3}=   \dfrac{\eta_{1}}{(1+\beta)\lambda^{2}(-81(18+\delta_{5})+243(1+\beta)^{2}(5+3\beta(4+3\beta))\lambda^{4}                                                                +54(1+\beta)^{4}(2+3\beta)\lambda^{8}+2(1+\beta)^{6}\lambda^{12})   }$
\\
\\
\\
\\
$\eta_{4}=   (-81(2\times 2^{1/3}\delta_{3}^{2/3}+18(18+\delta_{5})+2^{2/3}\delta_{3}^{1/3}(18+\delta_{5})) + 9(1+\beta)^{2}(243\beta^{2}(18+2^{2/3}\delta_{4}^{1/3})+
\\
\\
~~~~~~~~
12\beta(486 +27\times 2^{2/3}\delta_{4}^{1/3}-2^{1/3}\delta_{4}^{2/3})+135\times 2^{2/3}\delta_{4}^{1/3}-8\times 2^{1/3}\delta_{4}^{2/3}+18(153+\delta_{5}))\lambda^{4}-2(1+\beta)^{4}
\\
\\
~~~~~~~~~~
(243+2187\beta^{2}-81\beta(-18+2^{2/3}\delta_{4}^{1/3})-54\times 2^{2/3}\delta_{4}^{1/3}+2^{1/3}\delta_{4}^{2/3}\lambda^{8}+2(1+\beta)^{6}(-90-162\beta+2^{2/3}\delta_{4}^{1/3})\lambda^{12}
\\
\\
~~~~~~~
-4(1+\beta)^{8}\lambda^{16})$
\\
\\
$\eta_{5}= -81(18+\delta_{5})+243(1+\beta)^{2} (5+3\beta(4+3\beta))\lambda^{4}+54(1+\beta)^{4}(2+3\beta)\lambda^{8}+2(1+\beta)^{6}\lambda^{12}$
\\
\\
$\eta_{6}= 1-\dfrac{\eta_{4}}{54\eta_{5}}$
\\
\\
$\eta_{7}= \dfrac{\eta_{4}}{(1+\beta)^{2}\eta_{5}\lambda^{2}} $
\\
\\
$\eta_{8}= \sqrt{3}\surd \left (\eta_{6} \Big{(}3(3+\beta)^{2}+ \dfrac{\eta^{2}_{3}}{12}+  \dfrac{\eta^{2}_{4}}{27\eta^{2}_{5}} -\dfrac{2}{9}(1+\beta)^{2}\sqrt{\eta_{2}}\left(-3-\beta +\dfrac{\eta_{3}\sqrt{\eta_{6}}}{18}\right)\sqrt{\eta_{7}}\lambda^{2}-\dfrac{2}{81}(1+\beta)^{4}\sqrt{\eta_{2}}\eta^{3/2}_{7}\lambda^{4}
\right. \\ \left. 
 \qquad ~
 -\dfrac{1}{9} \eta_{3}(27\sqrt{\eta_{6}}+9\beta\sqrt{\eta_{6}}+4(1+\beta)\lambda^{2}+\dfrac{\eta_{4}(-12(3+\beta)+ \dfrac{1}{18} \eta_{3}(36\sqrt{\eta_{6}}-\dfrac{\eta_{1}}{2(1+\beta)\eta_{5}\lambda^{2}}+4(1+\beta)\lambda^{2}))   }{18\eta_{5}}  \Big{)}    \right)$
\\
\\
\\
$\eta_{9}=  81(324+18\times 2^{2/3}\delta^{1/3}_{3}+2\times 2^{1/3}\delta^{2/3}_{3}+18\delta_{5}+2^{2/3}\delta^{1/3}_{3}\delta_{5})-9(1+\beta)^{2}(2754+243\beta^{2}(18+2^{2/3}\delta^{1/3}_{4}) 
+\\ ~~~~~~~~~
12\beta(486+27\times2^{2/3}\delta^{1/3}_{4}-2^{1/3}\delta^{2/3}_{4})+135\times 2^{2/3}\delta^{1/3}_{4} -8\times 2^{1/3}\delta^{2/3}_{4}+18\delta_{5})\lambda^{4}+2(1+\beta)^{4}(243+2187\beta^{2}
-\\ ~~~~~~~~~ 
81\beta(-18+2^{2/3}\delta^{1/3}_{4})-54\times 2^{2/3}\delta^{1/3}_{4}+2^{1/3}\delta^{2/3}_{4})\lambda^{8}-2(1+\beta)^{6}(-90-162\beta+2^{2/3}\delta^{1/3}_{4})\lambda^{12}+4(1+\beta)^{8}\lambda^{16} $
\\
\\
\\
$\eta_{10}= \dfrac{\eta_{9}}{(1+\beta)^{2}\eta_{5}\lambda^{2}}$
\\

In the case of a tachyon field, we consider only the coupling $Q=\beta \dot{\rho_{\phi}}$. The point (3) shows the negativity of the eigenvalues under given conditions. Hence, it is a stable point. The phase portrait is shown in Fig. \ref{figtech}.
\section{Conclusions}
\label{sec:conc}
We investigated the interaction of a phantom field with a dark matter component in a spatially flat FLRW Universe. The choices of the coupling $Q$ in the conservation  equations were phenomenological and heuristic as there is no fundamental theory  of coupling strength in the dark sector was involved. We examined three different couplings, and studied the corresponding dynamical behavior and phase space. We paid attention on the stable point which could give rise to an accelerating phase. For all the three different couplings, we found $\Omega_{\phi}$, $w_{eff}$ and $w_{\phi}$. Our primary goal was to see if there exist late time scaling attractor with an accelerating phase and having the property $\Omega_{DE}/\Omega_{DM} \simeq {\cal{O}}(1)$. For the coupling $Q=\alpha \dot{\rho_m}$, we focused on Cases (3) and (5). In both cases the stationary points have an accelerating phase, but one of the stationary point  is stable and the other is a saddle point. In case  (3) the point is stable for $\alpha>1$ and $\lambda> \sqrt{3/ (\alpha-1)}$, and behaves as an attractive node. In this case, we obtained a stable fixed point with an accelerating Universe ($w_{eff}<-1/3$), however it corresponds to the case where dark energy completely dominates, as now we have  $\Omega_{\phi}=1$. Therefore, it does not solve the coincidence problem. The results  are shown in Fig. \ref{figint1a} and Fig. \ref{figint1b}. In the case of the coupling $Q=\beta \dot{\rho_{\phi}}$, we concentrated on Cases (2) and (4), and in  both cases the points   are stable but possessing non-accelerating phases as now $w_{\phi}$ is always positive (see Table \ref{tab2}). In the case (4), we considered two sets of the parameters as $\beta=-2.5$, $1 \leq \lambda < 1.5$ and $\beta=-2.5$, $ \lambda > 1.5$. In the first set, the point in Case (4) behaves as a stable point and give rise to a non-accelerating Universe ($w_{\phi}$ always positive). In the second set, it acts as a saddle point and has an accelerating Universe ($w_{\phi}<-1$). Thus, it can not solve coincidence problem either. The phase portrait, evolution of $w_{\phi}$ and $\Omega_{\phi}$ are displayed in Figs. \ref{figint2a} and \ref{figint2b}. For the linear combination of the coupling  $Q=\sigma (\dot{\rho}_{m}+\dot{\rho_{\phi}})$, we noticed that the stationary point in Case (3) is stable for $\sigma^2>1$ and $\lambda>0$,  but could not give rise to an accelerating Universe  as the equation of state is always positive. The phase portrait for this case is shown in Fig. \ref{figint3}, and it acts as an attractive node. 

For all the couplings considered here, our analysis showed that the coincidence problem cannot be alleviated in the coupled phantom field models. Similar results were also shown in \cite{saridakisa} for different couplings.

We also studied the dynamical behavior and stabilities for the coupled tachyon field with the coupling  $Q=\beta \dot{\rho_{\phi}}$. In this case, the eigenvalues of  the stationary point in Case (3) are negative. Therefore, it is a stable point.
\section*{Acknowledgements}
S.Li acknowledges to SDU-TH-2017001 and financial support by the NSFC Grant No. 11635009. A.W. is supported in part by NNSFC Grants Nos. 11375153 and 11675145, China.
\section*{Appendix}
For the sake of simplicity, we investigate the system of two first order differential equations, but it can be carried to a system of any number of equations. We study the following coupled differential equations for the variables $x(t)$ and $y(t)$ as
\begin{eqnarray}
\label{eqdotx}
\dot{x}&=&f(x,y,t)\\
\dot{y}&=&g(x,y,t)
\label{eqdoty}
\end{eqnarray}
where $f$ and $g$ are the functions of $x$, $y$ and $t$. If the functions $f$ and $g$ do not have explicit time-dependence then above equations are said to be an autonomous system. The dynamical analysis of the autonomous system can be investigated as following.

We can find the fixed or critical points by putting the left-hand-side of the autonomous system to zero. In other words, a point $(x_c, y_c)$ is said to be a critical point when it satisfies the following condition,
\begin{eqnarray}
f(x,y) \Big{\vert}_{(x_c,y_c)}&=&0\\
g(x,y)\Big{\vert}_{(x_c,y_c)}&=&0
\end{eqnarray}
The point $(x_c, y_c)$ would behave as an attractor when it meets the following condition,
\begin{eqnarray}
\Big{(}x(t), y(t) \Big{)}~ \longrightarrow ~(x_c,y_c) ~~~\text{for}~~~ t~ \longrightarrow ~ \infty
\end{eqnarray}
Next, we shall discuss the stability around the critical point. For this, we consider small perturbations $\delta x$ and $\delta y$ near the critical point as
\begin{eqnarray}
\label{eqdelx}
x&=&x_c+ \delta x\\
y&=&y_c+ \delta y
\label{eqdely}
\end{eqnarray}  
On putting equations (\ref{eqdelx}) and (\ref{eqdely}) into equations (\ref{eqdotx}) and (\ref{eqdoty}), we get first order differential equations,
\[ \frac{d}{dN} \left( \begin{array}{c} \delta x \\
\delta y \end{array} \right) = M \left( \begin{array}{c}
\delta x \\ \delta y \end{array} \right)\] 
where $N = ln (a)$ and matrix $M$ depends upon critical point $(x_c, y_c)$, and is written as
\[ M =  \left( \begin{array}{cc}
\frac{\partial f}{\partial x} & \frac{\partial f}{\partial y} \\ 
\frac{\partial g}{\partial x} & \frac{\partial g}{\partial y} \end{array} \right)_{(x=x_c, y=y_c)}\]
It contains two eigenvalues $\mu_1$, $\mu_2$, and the general solution for $\delta x$ and $\delta y$ is given as
\begin{eqnarray}
\delta x&=&k_1 e^{\mu_1 N}+ k_2 e^{\mu_2 N}\\
\delta y&=&k_3 e^{\mu_1 N}+ k_4 e^{\mu_2 N}
\end{eqnarray}  
where $k_1$, $k_2$, $k_3$ and $k_4$ are integration constants. Thus the sign of the  eigenvalues tell us the stability of the fixed points. Usually, the following classifications is used \cite{dw, tapan}:\\
(a) $\mu_1 < 0$ and $\mu_2 < 0$ $\longrightarrow$ Stable node\\
(b) $\mu_1 > 0$ and $\mu_2 > 0$ $\longrightarrow$ Unstable node\\
(c) $\mu_1 < 0$ and $\mu_2 > 0$~~or ($\mu_1 > 0$ and $\mu_2 < 0$) $\longrightarrow$ Saddle point\\
(d) The real parts of $\mu_1$ and $\mu_2$ are negative and the determinant of matrix $M$ is negative $\longrightarrow$ Stable spiral.

In case of (a) and (d), the fixed point is an attractor whereas in case of (b) and (c), it is not.

\end{document}